\documentclass[prd,a4paper,twocolumn,footinbib]{revtex4}
\usepackage[a4paper, hdivide={1.91cm,,1.165cm}, vdivide={1.83cm,,3.4cm}]{geometry}

\usepackage{amsmath,amssymb}
\usepackage{graphicx}
\usepackage{units}
\usepackage[hyperfootnotes=false,colorlinks,citecolor=blue]{hyperref}

\newcommand{\beq}{\begin{equation}}
\newcommand{\eeq}{\end{equation}}
\newcommand{\bea}{\begin{eqnarray}}
\newcommand{\ena}{\end{eqnarray}}

\def \L {\mathcal{L}} 

\def \epsilon {\varepsilon} 

\newcommand{\hc}{\ensuremath{\text{h.c.}}}
\newcommand{\BR}{\ensuremath{\text{BR}}}
\newcommand{\matrixx}[1]{\begin{pmatrix} #1 \end{pmatrix}} 

\allowdisplaybreaks

\begin{document}

\title{Zee-model predictions for lepton flavor violation}

\author{Julian Heeck}
\email[E-mail: ]{heeck@virginia.edu}
\thanks{ORCID:~\href{https://orcid.org/0000-0003-2653-5962}{0000-0003-2653-5962}.}
\affiliation{Department of Physics, University of Virginia,
Charlottesville, Virginia 22904-4714, USA}

\author{Anil Thapa}
\email[E-mail: ]{wtd8kz@virginia.edu}
\thanks{ORCID:~\href{https://orcid.org/0000-0003-4471-2336}{0000-0003-4471-2336}.}
\affiliation{Department of Physics, University of Virginia,
Charlottesville, Virginia 22904-4714, USA}

\hypersetup{
pdftitle={Zee-model predictions for lepton flavor violation},   
pdfauthor={Julian Heeck, Anil Thapa}
}


\begin{abstract}
The Zee model provides a simple model for one-loop Majorana neutrino masses. The new scalars can furthermore explain the long-standing deviation in the muon's magnetic moment and the recent CDF measurement of the $W$-boson mass. Together, these observations yield predictions for lepton flavor violating processes  that are almost entirely testable in the near future. The remaining parameter space makes testable predictions for neutrino masses.
\end{abstract}

\maketitle


\section{Introduction}

Neutrino oscillations have long proven that neutrinos are massive particles and that the individual lepton numbers $L_{e,\mu,\tau}$ are violated in nature. This by itself unavoidably induces \emph{charged}-lepton flavor violation (LFV),  but is unfortunately suppressed by powers of the minuscule neutrino mass $M^\nu$~\cite{Davidson:2022jai}. With the possible exception of neutrinoless double-beta decay ($0\nu\beta\beta$)~\cite{Rodejohann:2011mu}, all such neutrino-mass induced LFV is rendered unobservable with currently imaginable technology.

However, many neutrino-mass models also induce LFV processes with amplitudes \emph{unsuppressed} by $M^\nu$, with rates potentially in the observable range. Definite predictions are  hindered by our lack of knowledge about the masses of the new particles and their couplings, typically not uniquely fixed by the neutrino masses.
Only by fixing the new masses and couplings by tying them to other observables beyond the Standard Model (SM), e.g. anomalies, dark matter, or baryogenesis, can we hope to obtain testable predictions for LFV that allow for model falsification and goal posts for experimental sensitivities.

In this article, we perform a study along these lines for the Zee model~\cite{Zee:1980ai,Zee:1985id}. Here, the  SM is extended by a second scalar $SU(2)_L$ doublet and a charged singlet scalar, which leads to one-loop Majorana neutrino masses. The loop suppression already forces the new masses to be smaller than in tree-level neutrino-mass models, but still hopelessly out of range of LFV experiments in the worst-case scenario.
If we demand the new scalars to explain the long-standing anomaly in the muon's magnetic moment, however, we generically expect testable LFV, as we will quantify below.

The anomalous magnetic moment of the muon, $a_\mu \equiv (g-2)_\mu/2$, is a precisely calculated quantity in the SM~\cite{Aoyama:2020ynm}, equally precisely measured at BNL~\cite{Muong-2:2006rrc} and Fermilab~\cite{Muong-2:2021ojo}. Experiment and theory deviate by $4.2\sigma$, strongly hinting at a required new-physics contribution
\begin{align}
\Delta a_\mu = (2.51\pm 0.59)\times 10^{-9}\,.
\label{eq:amu}
\end{align}
Despite its existence for well over a decade~\cite{Muong-2:2006rrc}, the status of this anomaly is not settled yet, with recent lattice-QCD measurements casting doubt on the SM prediction or at least its uncertainty~\cite{Borsanyi:2020mff}.
With no consensus yet in the community on this issue, we will take the deviation in Eq.~\eqref{eq:amu} at face value and explore its resolution within the Zee model.

An even more significant deviation from an SM prediction was recently reported by the CDF collaboration~\cite{CDF:2022hxs} in their legacy measurement of the $W$-boson mass:
\begin{align}
M_W^\text{CDF} = \unit[(80.4335\pm 0.0094)]{GeV}\,.
\label{eq:CDF}
\end{align}
This exceeds the similarly-precise SM prediction~\cite{Awramik:2003rn} by an astonishing $7\sigma$ and has led to a flurry of activity regarding possible resolutions, including appeals to new physics.
The Zee model under consideration here is capable of ameliorating this CDF anomaly~\cite{Chowdhury:2022moc,Primulando:2022vip} and we will study the relevant parameter space below.

In the remainder of this article we will show how the Zee-model explanation of $a_\mu$ and $M_W$ leads to predictions for LFV and neutrino observables. 
We start by reviewing the Zee model in Sec.~\ref{sec:model} and introduce relevant observables and our parametrization in Sec.~\ref{sec:observables}. Our numerical scan of the parameter space is introduced in Sec.~\ref{sec:numerical_scan} and we discuss our finding in Sec.~\ref{sec:discussion}. We conclude in Sec.~\ref{sec:conclusion}.

\section{Zee Model}
\label{sec:model}

The Zee model extends the SM by a charged scalar $\eta^+$ and a second Higgs doublet $H_2$, with the following relevant interaction terms
in the Lagrangian
\begin{align}
\L = - \bar{L}^c f L  \eta^+ - \bar{\ell}\tilde{Y} L \tilde{H}_1- \bar{\ell} Y L \tilde{H}_2 + \mu H_1 H_2 \eta^-+\hc ,
\label{eq:lag}
\end{align}
suppressing flavor and $SU(2)_L$ indices. 
Without loss of generality we can rotate the two scalar doublets to the Higgs basis~\cite{Georgi:1978ri}, so that only $H_1$  acquires a vacuum expectation value, $\langle H_1\rangle 
\equiv v/\sqrt{2} \simeq \unit[174]{GeV}$. $M_\ell = \tilde{Y} v/\sqrt{2}$ is the charged lepton mass matrix, chosen to be diagonal without loss of generality. A similar coupling of $H_1$ to quarks yields quark masses and mixing, whereas we neglect the $H_2$ coupling to quarks in order to simplify the analysis below.
For further simplification, we ignore mixing between the CP-even neutral scalars in $H_1$ and $H_2$, i.e.~work in the alignment limit. 
The $\mu$ term in the Lagrangian will induce a mixing of $\eta^+$ with the charged scalar contained in $H_2$; we denote the mixing angle by $\phi$ and the two mass eigenstates by $h^+$ and $H^+$, see Ref.~\cite{Barman:2021xeq} for details.
Finally, $Y$ is an arbitrary complex Yukawa matrix while $f$ is antisymmetric in flavor space.

\begin{figure}
    \centering
    \includegraphics[width=0.3\textwidth]{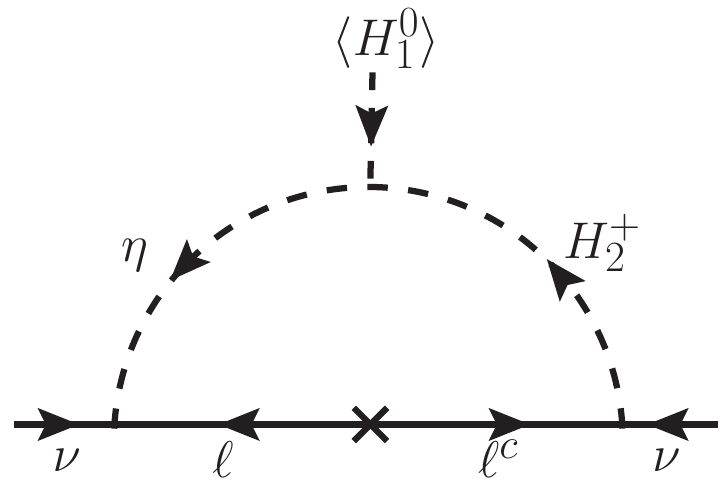}
    \caption{Radiative neutrino mass diagram in the Zee model.}
    \label{fig:numass}
\end{figure}

The simultaneous presence of $f$, $Y$, and $\mu$ breaks lepton number by two units and leads to Majorana neutrino masses at one-loop level through the diagrams in Fig.~\ref{fig:numass}:
\begin{align}
M^\nu = \kappa \left( f M_\ell Y + Y^T M_\ell f^T\right),
\label{eq:Mnu}
\end{align}
with $\kappa\equiv (16\pi^2)^{-1} \sin 2\phi \log(m_{h^+}^2/m_{H^+}^2)$.
This Zee-model expression  does not impose any constraints on the form of $M^\nu$, i.e.~does not make predictions about mixing angles or masses.
However, as we will show below, viable $M^\nu$ textures unavoidably lead to LFV amplitudes unsuppressed by neutrino masses. These arise from the couplings $Y$ and $f$, mediated by the new scalars~\cite{Herrero-Garcia:2017xdu}.

\section{LFV and other observables}
\label{sec:observables}

\begin{figure}
    \centering
    \includegraphics[width=0.33\textwidth]{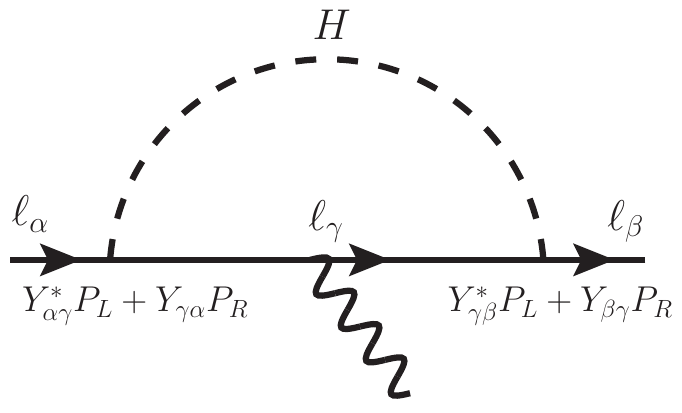}
    \includegraphics[width=0.28\textwidth]{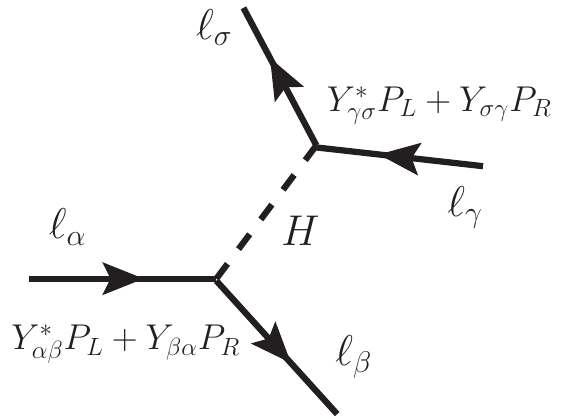}
    \caption{Top: Feynman diagram leading to LFV $\ell_\alpha \to \ell_\beta \gamma$ via the neutral scalar $H$ (or $A$). The same diagram also leads to $(g-2)_\ell$ for $\alpha=\beta=\ell$.\\
     Bottom: Diagram for tree-level trilepton decay $\ell_\alpha \to \ell_\beta \bar{\ell}_\gamma \ell_\sigma$. 
    }
    \label{fig:LFV}
\end{figure}

Expressions for LFV rates within the Zee model have long been derived in the literature \cite{Lavoura:2003xp, He:2011hs, Herrero-Garcia:2017xdu, Cai:2017jrq, Crivellin:2015hha}.
At tree level, these include the trilepton decays $\ell_\alpha \to \ell_\beta \bar{\ell}_\gamma \ell_\sigma$, illustrated in Fig.~\ref{fig:LFV}.
Current limits and expected near-future sensitivities are collected in Tab.~\ref{tab:201}.
At loop level, we find dipole operators $\ell_\alpha \ell_\beta \gamma$ that include the desired magnetic moment of the muon but also \emph{electric} dipole moments (EDMs) of muon and electron as well as LFV amplitudes for $\mu\to e\gamma $ and others.
In addition to the one-loop diagrams of Fig.~\ref{fig:LFV}, we also include two-loop Barr--Zee~\cite{Barr:1990vd,Bjorken:1977vt} contributions. The most relevant contributions arise from a photon propagator with neutral scalars and charged lepton loop~\cite{Ilisie:2015tra, Crivellin:2015hha, Cherchiglia:2016eui, Cherchiglia:2017uwv,Frank:2020smf}. Various other diagrams involving $Z$ boson, charged scalars instead of lepton loop, and diagrams involving $W^+W^-H_1$ and $H_1 H_2 H_2$ are not considered here as they are suppressed. For instance, the contribution from $W^+W^-H_1$ and $H_1 H_2 H_2$ vanish in the alignment limit, and the contribution from charged scalar in the loop can be made small by taking the relevant quartic coupling small.

\begin{table}[t!]
\begin{center}
\begin{tabular}{|c|c|c|}\hline
& Present bound & Future sensitivity \\\hline\hline

$\mu\to e \gamma$&$4.2\times 10^{-13}$~\cite{MEG:2016leq} &$6\times 10^{-14}$~\cite{Baldini:2013ke,Meucci:2022qbh}  \\ \hline
$\tau\to e \gamma$&$3.3\times 10^{-8}$~\cite{BaBar:2009hkt} &$9 \times 10^{-9}$~\cite{Belle-II:2022cgf}  \\ \hline
$\tau\to \mu \gamma$&$4.4\times 10^{-8}$~\cite{BaBar:2009hkt} &$7 \times 10^{-9}$~\cite{Belle-II:2022cgf}  \\ \hline 
$\mu\to eee$&$1.0\times 10^{-12}$~\cite{SINDRUM:1987nra} &$\sim 10^{-16}$~\cite{Blondel:2013ia,Mu3e:2020gyw}  \\ \hline
$\tau\to eee$&$2.7\times 10^{-8}$~\cite{Hayasaka:2010np} &$5 \times 10^{-10}$~\cite{Belle-II:2022cgf}  \\ \hline
$\tau\to \mu\mu\mu$&$2.1\times 10^{-8}$~\cite{Hayasaka:2010np} &$3.5 \times 10^{-10}$~\cite{Belle-II:2022cgf}  \\ \hline
$\tau^-\to e^-\mu^+\mu^-$&$2.7\times 10^{-8}$~\cite{Hayasaka:2010np} &$4.5 \times 10^{-9}$~\cite{Belle-II:2022cgf}  \\ \hline
$\tau^-\to \mu^-e^+e^-$&$1.8\times 10^{-8}$~\cite{Hayasaka:2010np} &$3 \times 10^{-10}$~\cite{Belle-II:2022cgf}  \\ \hline
$\tau^-\to e^+\mu^-\mu^-$&$1.7\times 10^{-8}$~\cite{Hayasaka:2010np} &$2.5 \times 10^{-10}$~\cite{Belle-II:2022cgf}  \\ \hline
$\tau^-\to \mu^+e^-e^-$&$1.5\times 10^{-8}$~\cite{Hayasaka:2010np} &$2.2 \times 10^{-10}$~\cite{Belle-II:2022cgf}  \\ \hline
$e^-\mu^+ \leftrightarrow e^+ \mu^-$ & $8.3 \times 10^{-11}$~\cite{Willmann:1998gd} &$2 \times 10^{-14}$~ \cite{Bai:2022sxq}\\ \hline
$\mu \leftrightarrow e$\hfill [Au]& $7 \times 10^{-13}$~\cite{SINDRUMII:2006dvw} & $-$ \\ 
~conv. \hfill [Al]& $-$ & $ 6\times 10^{-17}$~\cite{Mu2e:2014fns,Mu2e-II:2022blh}\\ \hline
$\mu$EDM & $1.9 \times 10^{-19}$~\cite{Muong-2:2008ebm} & $6 \times 10^{-23}$~\cite{Adelmann:2021udj,Sato:2021aor} \\ \hline
$e$EDM & $1.1 \times 10^{-29}$~\cite{ACME:2018yjb} & $\sim 10^{-30}$~\cite{Hiramoto:2022fyg,Roussy:2022cmp} \\ \hline
$\Delta a_e^{\rm comb}$ & $(2.8 \pm 2.9) \times 10^{-13}$ & $-$\\ \hline
\end{tabular}
\end{center}
\caption{Current experimental bounds on $\BR(\ell_i\to \ell_j\gamma)$, $\BR(\ell_i\to \ell_k\ell_m\ell_n)$,  muonium--antimuonium conversion $P(e^-\mu^+ \leftrightarrow e^+ \mu^-)$, $\mu\to e$ conversion in nuclei, muon and electron EDM.
All bounds are at $90\%$ C.L.~except for $\mu$EDM, which is at $95\%$ C.L. 
Future sensitivities are given in the last column. 
There is disagreement between experiments for the anomalous magnetic moment of the electron $a_e$:  $(-88 \pm 36)\times 10^{-14}$~\cite{Parker:2018vye} and $(48\pm 30)\times 10^{-14}$~\cite{Morel:2020dww}; we use the weighted combination shown in the last row.
}
\label{tab:201}
\end{table}

In the alignment limit and without $H_2$ couplings to quarks, $\mu\to e $ conversion in nuclei only arises through the same dipole operator that generates $\ell\to\ell' \gamma$.
As such, we find the relation for conversion in aluminum (as relevant for the upcoming COMET and Mu2e):
\begin{align}
\BR (\mu\to e, \text{Al}) \simeq 0.0027 \ \BR(\mu\to e\gamma)\,,
\end{align}
exhibiting the expected suppression by $\alpha\sim 1/137$~\cite{Kitano:2002mt,Heeck:2022wer}.
Currently, the best limits on the $\mu$--$e$ dipole operator come from $\mu\to e\gamma$ in MEG~\cite{MEG:2016leq}, but will eventually be superseded by $\mu$-to-$e$ conversion in Mu2e~\cite{Mu2e:2014fns,Mu2e-II:2022blh}.

Interestingly, the Zee model is not only constrained through $\Delta |L_\alpha| =1$ LFV processes, but can also generate testable rates for the $|\Delta L_\mu| =|\Delta L_e|=2$ process of muonium ($M = e^-\mu^+$)  to  antimuonium ($\bar{M}= e^+\mu^-$) conversion~\cite{Pontecorvo:1957cp,  Jentschura:1997tv, Clark:2003tv, Fukuyama:2021iyw}.
The oscillation probability was constrained by PSI to $P(M\leftrightarrow\bar{M}) < 8.3 \times 10^{-11}$ at $90\%$ C.L. \cite{Willmann:1998gd}, while a sensitivity at the level of $\mathcal{O}(10^{-14})$ is expected in the future~\cite{Bai:2022sxq}. These oscillations place a stringent constraint on the Yukawa couplings $Y_{e\mu}$ and $Y_{\mu e}$. 
The oscillation probability is given by~\cite{Conlin:2020veq, Fukuyama:2021iyw}
\begin{equation}
    P(M\to \bar{M}) \simeq \frac{64 \alpha^6 m_e^6 \tau_\mu^2}{\pi^2} G_{M\bar{M}}^2 \,,
    \label{eq:probmm}
\end{equation}
with muon lifetime $\tau_\mu$ and  Wilson coefficient
\begin{equation}
    G_{M\bar{M}}^2 \simeq 0.32\, \left| \frac{3 G_3}{2} + \frac{G_{45}}{4}  \right|^2 + 0.13\, \left| \frac{G_{45}}{4}-0.68\, G_3  \right|^2 ,
\end{equation}
with the following coefficients in the alignment limit:~\cite{Fukuyama:2021iyw}
\begin{align}
    G_{45} &\equiv - \frac{Y_{e\mu}^{*2} + Y_{\mu e}^2}{8 \sqrt{2}} \left( \frac{1}{m_H^2} - \frac{1}{m_A^2} \right) , \\
    G_{3} &\equiv - \frac{Y_{e\mu}^{*}  Y_{\mu e}} {8 \sqrt{2}} \left( \frac{1}{m_H^2} + \frac{1}{m_A^2} \right)  . 
\end{align}

Finally, the mass splitting within the $SU(2)_L$ doublet $H_2$ breaks the SM's custodial symmetry and thus changes the relationship between $W$ and $Z$ boson mass. This can be used to accommodate the CDF anomaly.
Since we restrict ourselves to masses above the electroweak scale, the relevant effects can be parameterized by the oblique parameters $S$ and $T$~\cite{Peskin:1990zt,Peskin:1991sw}, which modify~\cite{Maksymyk:1993zm}
\begin{align}
        M_W \simeq M_W^{\rm SM} \left[1 - \frac{\alpha (S-2 c^2_W T)}{4(c_W^2-s^2_W)}  \right] ,  
\label{eq:MW-STU}
\end{align}
with $s_W\equiv \sin \theta_W$ and $c_W\equiv \cos \theta_W$.
Matching the CDF value from Eq.~\eqref{eq:CDF} fixes one linear combination of $S$ and $T$; the orthogonal combination is constrained from other electroweak data.
Numerous global fits have been performed following the wake of the CDF result to identify the preferred region of $S$ and $T$, here we will use the results of Ref.~\cite{Asadi:2022xiy}, both for the $2\sigma$ regions that explain CDF and those obtained by ignoring the CDF result, dubbed PDG. This allows us to see the impact of the CDF result on the Zee-model parameter space.
Similar results have been obtained in other fits~\cite{Lu:2022bgw}.
The Zee-model expression for $S$ and $T$ can be found in Refs.~\cite{Haber:2010bw,Herrero-Garcia:2017xdu} and will not be displayed here.

\subsection{Limiting Cases without LFV}
\label{sec:limiting_cases}

Before delving into the most general case, let us study limiting cases of coupling structures that lead to heavily suppressed or even vanishing LFV. First off, let us assume that $H_2$ is much heavier than $\eta^+$ or $|Y|\ll |f|$. In that case, $\eta^+$ will induce the dominant LFV, except for the following textures:
\begin{itemize}
\item \textbf{TX-F23:} Setting $f_{e\mu}=f_{e\tau}=0$ eliminates LFV through $\eta^+$ -- because we can assign $L_\tau (\eta^+) = L_\mu (\eta^+) =-1$ --  and predicts the one-zero texture $M^\nu_{ee}=0$, i.e.~no $0\nu\beta\beta$. However, this requires a specific $Y$  with little freedom to evade LFV through $H_2$ except by pushing its mass to high values.
\item \textbf{TX-F13:} Similarly, setting $f_{e\mu}=f_{\mu\tau}=0$ eliminates $\eta^+$ LFV and generates $M^\nu_{\mu\mu}=0$. 
\item \textbf{TX-F12:} Lastly, setting $f_{e\tau}=f_{\mu\tau}=0$ eliminates $\eta^+$ LFV and generates $M^\nu_{\tau\tau}=0$. 
\end{itemize}
Additional constraints on $\eta^+$ can be found in Ref.~\cite{Crivellin:2020klg}.
Alas, the $\eta^+$ contribution to $a_\mu$ is unavoidably of the wrong sign, rendering these three cases impotent to obtain Eq.~\eqref{eq:amu}. Only the scalars within $H_2$ can generate the desired sign for $a_\mu$ and can therefore not be pushed to arbitrarily high values. Since the three cases above allow for very little freedom in the $Y$ entries, a light $H_2$ might explain $a_\mu$ but will generate far too large LFV.

A more useful starting point is $|f|\ll |Y|$ or $H_2$ lighter than $\eta^+$. In this case, $a_\mu$ can have the correct sign and LFV will be dominated by $H_2$ through $Y$, see Fig.~\ref{fig:LFV}. Once again we can identify textures that suppress LFV:
\begin{itemize}
\item Diagonal $Y$: this case is by now excluded because it leads to an $M^\nu$ with three texture zeros, incompatible with oscillation data~\cite{Xing:2004ik}. We therefore unavoidably have \emph{off-diagonal} entries in $Y$!
\item $\Delta L_\alpha=1$ LFV decays can be evaded by choosing $Y$ to be of the forms
\begin{align}
Y_{E_3} &= \matrixx{ 0 & 0 & 0 \\ 0 & 0 & Y_{\mu\tau}\\ 0 & Y_{\tau\mu} & 0} ,\\ 
Y_{B_1} &= \matrixx{ 0 & 0 & Y_{e\tau} \\ 0 & 0 & 0\\ Y_{\tau e} & 0 & 0} , \label{eq:TXB1} \\ 
Y_{B_2} &= \matrixx{ 0 & Y_{e\mu} & 0 \\ Y_{\mu e} & 0 & 0\\ 0 & 0 & Y_{\tau \tau}} , \label{eq:TXB2}
\end{align}
which give rise to the $M^\nu$ two-zero textures~\cite{Frampton:2002yf} $E_3$, $B_1$, and $B_2$, respectively. 
The first of these is not compatible with oscillation data and thus requires additional entries in $Y$, which unavoidably generate LFV decays.
$Y_{B_1}$ can lead to a viable $M^\nu$ but does not contain any muon couplings to resolve $(g-2)_\mu$.
$Y_{B_2}$ on the other hand is compatible with oscillation data and  has muon couplings that can explain $(g-2)_\mu$. However, despite $\Delta L_\alpha=1$ lepton decays being absent for this texture, $Y_{B_2}$ does induce $\Delta L_\alpha=2$ muonium--antimuonium conversion as well as electron dipole couplings that render it utterly insufficient to explain $(g-2)_\mu$, essentially because the relevant couplings are linked by $Y_{e\mu}\sim 70\,Y_{\mu e}$, 
see the appendix for details.
\end{itemize}
From the above limiting cases we must conclude that any texture of $Y$ that explains $(g-2)_\mu$ and gives valid neutrino parameters has entries that lead to  LFV. As we will see below, the required electroweak-scale scalars to explain $(g-2)_\mu$ make it nearly impossible to suppress said LFV arbitrarily and actually make most of the model testable with near-future LFV experiments.

\subsection{General Parametrization}

In order to efficiently study the Zee-model parameter space, we use the parametrization from Ref.~\cite{Machado:2017flo} to solve Eq.~\eqref{eq:Mnu} for the Yukawa matrix $Y$ as
\begin{align}
Y &= \kappa^{-1} M_\ell^{-1} (Z + Q)\,, \label{eq:Yuk}\\
Z &\equiv \matrixx{ 
-\frac{M^\nu_{e\tau}}{f_{e\tau}} & 0 & -\frac{M^\nu_{\tau\tau}}{2f_{e\tau}}\\
0 & \frac{f_{e\mu} M^\nu_{\tau\tau}-2f_{e\tau} M^\nu_{\mu\tau}}{2 f_{e\tau}f_{\mu\tau}} & 0\\
\frac{M^\nu_{ee}}{2 f_{e\tau}} & \frac{M^\nu_{\mu\mu}}{2 f_{\mu\tau}} & 0} ,\\
\nonumber\\  
Q &\equiv \matrixx{
 2 q_4-\frac{f_{\mu\tau} }{f_{e\tau}} q_1 & \frac{f_{\mu\tau}}{f_{e\tau}} (q_4-q_2) & -\frac{2 f_{\mu\tau}}{f_{e\mu}} q_4-\frac{f_{\mu\tau}}{f_{e\tau}} q_3 \\
 q_1 & q_2+q_4 & \frac{2 f_{e\tau}}{f_{e\mu}} q_4+q_3 \\
 -\frac{f_{e\mu}}{f_{e\tau}}  q_1 & \frac{f_{e\mu}}{f_{e\tau}} (q_4-q_2) & -\frac{f_{e\mu}}{f_{e\tau}}  q_3 
}.
\end{align}
assuming the three (complex) entries of $f$ to be nonzero; one entry of $f$ is fixed by the constraint equation
\begin{align}
\begin{split}
0 &=f_{\mu\tau}^2 M^\nu_{ee} - 2 f_{e\tau} f_{\mu\tau} M^\nu_{e\mu} + 2 f_{e\mu} f_{\mu\tau} M^\nu_{e\tau} \\
&\quad+ f_{e\tau}^2 M^\nu_{\mu\mu} - 2 f_{e\mu} f_{e\tau} M^\nu_{\mu\tau} + f_{e\mu}^2 M^\nu_{\tau\tau}\,.
\end{split}
\label{eq:constraint}
\end{align}
$Q$ drops out of the neutrino mass formula and contains four complex parameters $q_j$.
It is straightforward to show that the so-defined $Y$ indeed satisfies the $M^\nu$ equation~\eqref{eq:Mnu} and contains the correct number of free parameters~\cite{Machado:2017flo}. 
This parametrization is convenient as it allows us to use the known neutrino parameters as input and is far simpler than other expressions put forward in the literature~\cite{Cordero-Carrion:2019qtu}.

\subsection{Muonphilic textures}
\label{sec:muonphilic_cases}

In Sec.~\ref{sec:limiting_cases} we have argued that a resolution of $a_\mu$ without LFV is impossible within the Zee model. The parametrization from above allows us to easily study textures that explain $a_\mu$ and still suppress LFV sufficiently.
We aim to find muonphilic Yukawa textures, i.e.~those with a large $Y_{\mu\mu}$ entry, as this will lead to a large $a_\mu$ contribution by the neutral scalars $A$ and $H$~\cite{Chowdhury:2022moc}. A large $Y_{\mu\mu}$ immediately requires highly suppressed $Y_{e\mu}$ and $Y_{\mu e}$ in order to suppress $\mu\to e\gamma$ and $\mu\to 3e$. This can be achieved via $q_1 = 0$ and $q_4=q_2$ in the general parametrization. 

The remaining $q_2$ and $q_3$ can be used to set two more entries of $Y$ to zero, e.g.~$Y_{ee} = Y_{e\tau} = 0$, leading to $ \kappa Y =$
\begin{align}
   \matrixx{ 0 & 0 & 0\\
    0 & \frac{2 f_{\mu\tau} M^\nu_{e\tau} - 2 f_{e\tau} M^\nu_{\mu\tau} + f_{e\mu} M^\nu_{\tau\tau}}{2 f_{e\tau} f_{\mu\tau} m_\mu} & -\frac{M^\nu_{\tau\tau}}{2 f_{\mu\tau} m_\mu}\\
    \frac{M^\nu_{ee}}{2f_{e\tau} m_\tau} & \frac{M^\nu_{\mu\mu}}{2 f_{\mu\tau} m_\tau} &  \frac{2 f_{\mu\tau} M^\nu_{e\tau} + f_{e\mu} M^\nu_{\tau\tau}}{2 f_{e\tau}f_{\mu\tau} m_\tau}} .
    \label{eq:muontx1}
\end{align}
Interestingly, the limit $M^\nu_{ee} \to 0$ leads to electron-number conservation, at least through the $Y$ interactions. This automatically eliminates all muonic LFV, which pose the most serious threat to an explanation of $a_\mu$. It is not sufficient though, as tauonic LFV is generically too large as well.
However, even the remaining off-diagonal entries of $Y$, which lead to the LFV decays $\tau\to 3\mu$ and $\tau\to \mu\gamma$, can be suppressed by taking $f_{e\tau} \ll f_{\mu\tau}$. In this limit, $Y_{\mu\mu}$ is the dominant entry, $Y_{\tau\tau} \simeq Y_{\mu\mu} m_\mu/m_\tau$ is the second-largest entry, and $Y_{\tau\mu,\mu\tau}$ are suppressed. 
For this particular texture, $a_\mu$ can be explained without testable LFV, even in future experiments. We stress that this relied on $M^\nu_{ee} = 0$, which constitutes a testable prediction in the neutrino sector: the absence of $0\nu\beta\beta$~\cite{Rodejohann:2011mu}, and normal hierarchy for the neutrino mass spectrum (see Tab.~\ref{tab:texture_zeros}).

\begin{table}[tb]
\begin{tabular}{c|c|c|c}
texture zero & ordering & $\sum_j m_j/\unit{meV}$ & $\langle m_{\beta\beta}\rangle/\unit{meV}$\\
\hline
$M_{ee} = 0$ & normal & $\in [60,65]$ & $0$\\
$M_{ee} = 0$ & inverted & -- & -- \\
$M_{\mu\mu} = 0$ & normal & $>150$ & $>41$\\
$M_{\mu\mu} = 0$ & inverted & $>98$ & $>15$\\
\end{tabular}
\caption{Predictions for the sum of neutrino masses $\sum_j m_j$ and the effective $0\nu\beta\beta$ Majorana neutrino mass $\langle m_{\beta\beta}\rangle$ from the texture zeros $M_{ee}=0$ and $M_{\mu\mu}=0$, using the $3\sigma$ ranges for the oscillation parameters from Ref.~\cite{Esteban:2020cvm}.
}
\label{tab:texture_zeros}
\end{table}

Instead of using $q_2$ and $q_3$ to eliminate $Y_{ee}$ and $Y_{e\tau}$, one can set $Y_{\mu\tau}=0$ via $q_3 = - 2 f_{e\tau} q_2/ f_{e\mu}$, which gives the  texture $\kappa Y =$
\begin{align}
   \matrixx{ \frac{-M^\nu_{e\tau}+2 q_2 f_{e\tau}}{m_e f_{e\tau}} & 0 &  -\frac{M^\nu_{\tau\tau}}{2 f_{e\tau} m_e}\\
    0 & \frac{-2 f_{e\tau} M^\nu_{\mu\tau} + f_{e\mu} M^\nu_{\tau\tau} + 4 f_{e\tau} f_{\mu\tau} q_2}{2 f_{e\tau} f_{\mu\tau} m_\mu} & 0\\
    \frac{M^\nu_{ee}}{2f_{e\tau} m_\tau} & \frac{M^\nu_{\mu\mu}}{2 f_{\mu\tau} m_\tau} &  \frac{2 q_2}{m_\tau}} .
    \label{eq:muontx2}
\end{align}
Here, dangerous muonic LFV can be evaded by requiring $M^\nu_{\mu\mu} = 0$, which leads to a muon-number conserving $Y$. Once again this would not be sufficient; tauonic LFV have to be suppressed via the hierarchy  $f_{\mu\tau}\ll f_{e\tau}$. $q_2$ has to be small as well, extreme cases include $q_2=0$ (which gives $Y_{\tau\tau}=0$) and $q_2=M^\nu_{e\tau}/2f_{e\tau}$ (which gives $Y_{ee}=0$).
The above texture makes it possible to explain $a_\mu$ while suppressing LFV below future sensitivities, but hinges on $M^\nu_{\mu\mu} = 0$, which is again a testable prediction in the neutrino sector, as shown in Tab.~\ref{tab:texture_zeros}.

\section{Numerical Analysis}
\label{sec:numerical_scan}

With all relevant observables at our disposal we can numerically explore the Zee-model parameter space that explains $a_\mu$ (and CDF) to find LFV predictions.
The parametrization from Eq.~\eqref{eq:Yuk} allows us to use neutrino data as an input; we take the $3\sigma$ ranges of the oscillation parameters from the global fit~\cite{Esteban:2020cvm}, distinguishing between normal and inverted ordering. 
As an upper bound on the absolute neutrino mass we use $\unit[0.8]{eV}$~\cite{KATRIN:2021uub}.

We scan over two $f_{ij} = [10^{-15}, \sqrt{4 \pi}]$ -- the third one being determined by Eq.~\eqref{eq:constraint} -- and $|q_i| = [10^{-25},\ {\rm Max} |q_i|]$, while keeping the phases arbitrary and demanding the Yukawa  couplings to remain perturbative. The conservative upper bounds from perturbativity for $\kappa>0$ are
\begin{align}
    |q_1| & < \sqrt{4\pi} m_\mu \kappa\,,\\
    |q_2| & < \sqrt{4\pi} \left|\frac{f_{e\tau}}{f_{\mu\tau}}\right| m_e \kappa+\sqrt{\pi} \left|\frac{f_{e\mu}}{f_{e\tau}}\right| m_\mu \kappa+\sqrt{\pi}  m_\tau \kappa\,,\\
    |q_3| & < \sqrt{4\pi} \left|\frac{f_{e\tau}}{f_{e\mu}}\right| m_\tau \kappa\,,\\
    |q_4| & < \sqrt{\pi} \left|\frac{f_{e\mu}}{f_{e\tau}}\right| m_\mu \kappa+\sqrt{\pi} m_\tau \kappa\,.
\end{align}
In addition to the two $f_{ij}$ and four complex parameters $q_j$, the model has the following parameters that characterize the LFV while correlating it with $M^\nu$ and $(g-2)_\mu$: 
\begin{equation}
    \{m_H, m_A, m_{h^+}, m_{H^+}, \phi \} \, .
    \label{eq:param}
\end{equation}
The charged-scalar masses are scanned over $[0.1,100]$ TeV. The mixing angle $\phi$ is parameterized by the mass-square difference $m_{h^+}^2-m_{H^+}^2$ and a cubic coupling $\mu$ (cf. Eq.~\eqref{eq:lag}), where we take $\mu$ up to a maximum value of about $4.1$ times the heavier charged scalar mass to be consistent with charge-breaking minima~\cite{Barroso:2005hc, Babu:2019mfe}. This leaves us with $\{ m_H, m_A \}$ which we numerically solve to obtain the desired $a_\mu$ and $\chi^2$ of CDF/PDG within $2\sigma$, which are the functions of parameters given in Eq.~\eqref{eq:param}. The resulting $\{ m_H, m_A \}$ are, of course, often unphysical.

The above scan automatically satisfies any neutrino-mass constraints and aims to explain the $a_\mu$ and CDF anomalies. However, most of these points in parameter space are already excluded by current LFV limits. In an effort to find corners of parameter space where LFV is suppressed, we also perturb around the previously identified textures that evade $\Delta L_\alpha = 1$ to make sure the procedure adopted is as unbiased as possible. Eventually, all scans are combined, resulting in $\mathcal{O}(10^{9})$ points.

\section{Discussion}
\label{sec:discussion}

\begin{figure}
    \centering
    \includegraphics[width=0.5\textwidth]{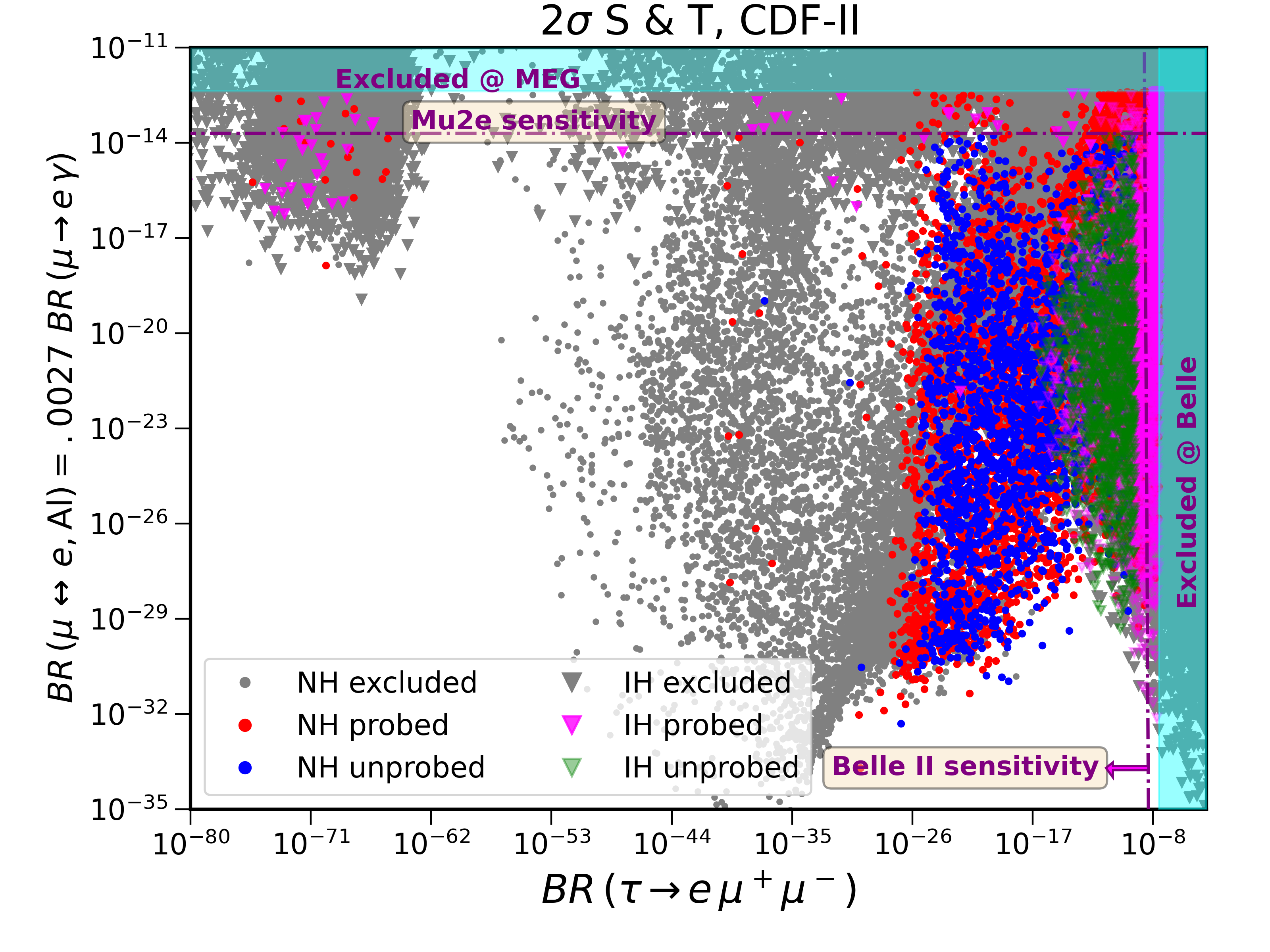}\\
    \includegraphics[width=0.5\textwidth]{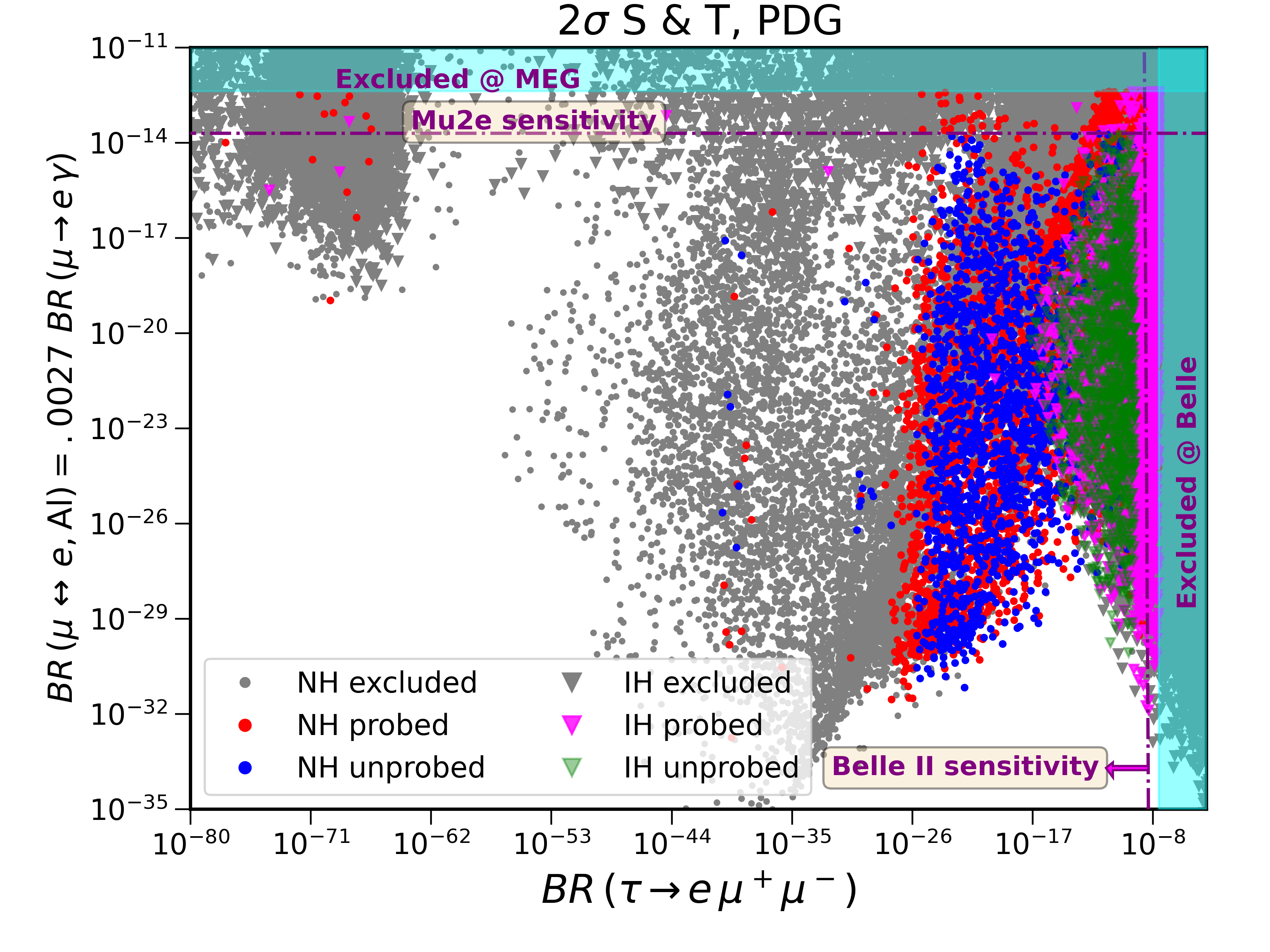}
    \caption{LFV observables $\tau \to e \mu^+ \mu^-$ and $\mu \to e \gamma$ for normal ($\bullet$) and inverted ordering ($\blacktriangledown$).
    All points explain $(g-2)_\mu$; in the top plot, we also explain the CDF anomaly, while we ignore CDF in the bottom plot.
     Gray data points are excluded by the current experimental bounds listed in Tab.~\ref{tab:201}. Red/pink data points can be probed in future experiments. Blue/green points cannot be probed in future experiments. Cyan colored band (dashed purple line) is the current exclusion (future sensitivity) limit for $\tau \to e \mu^+ \mu^-$ and $\mu \to e \gamma$.}
    \label{fig:prediction}
\end{figure}

In Fig.~\ref{fig:prediction}, we show some relevant observables,  $\tau \to e \mu^+ \mu^-$ and $\mu \to e \gamma$,  that can probe a lot of the parameter space and convey the qualitative results of our numerical scan. All points resolve the $a_\mu$ anomaly, give valid neutrino parameters, and have perturbative Yukawas and scalar masses above or around the electroweak scale.
In the top figure of Fig.~\ref{fig:prediction}, all points furthermore explain the CDF anomaly within $2\sigma$, while in the bottom plot the CDF anomaly is ignored and we satisfy the PDG results for $S$ and $T$.

The gray data points, which make up the vast majority of our scan, are already excluded by the current experimental bounds listed in Tab.~\ref{tab:201}. 
Red and pink data points are currently valid and can be probed in future experiments (defined through the last column in Tab.~\ref{tab:201}), and correspond to different neutrino mass orderings. These include the textures recently put forward in Refs.~\cite{Chowdhury:2022moc,Primulando:2022vip}.
Finally, blue and green points lead to LFV that is suppressed beyond near-future sensitivities; these points are nearly impossible to find in an unbiased scan and all correspond to perturbations of the two textures~\eqref{eq:muontx1} and~\eqref{eq:muontx2}. The reader should not be led astray by their seemingly large number and density in Fig.~\ref{fig:prediction}, these points correspond to a tiny region in parameter space that we sampled very thoroughly.

\begin{figure}
    \centering
    \includegraphics[width=0.49\textwidth]{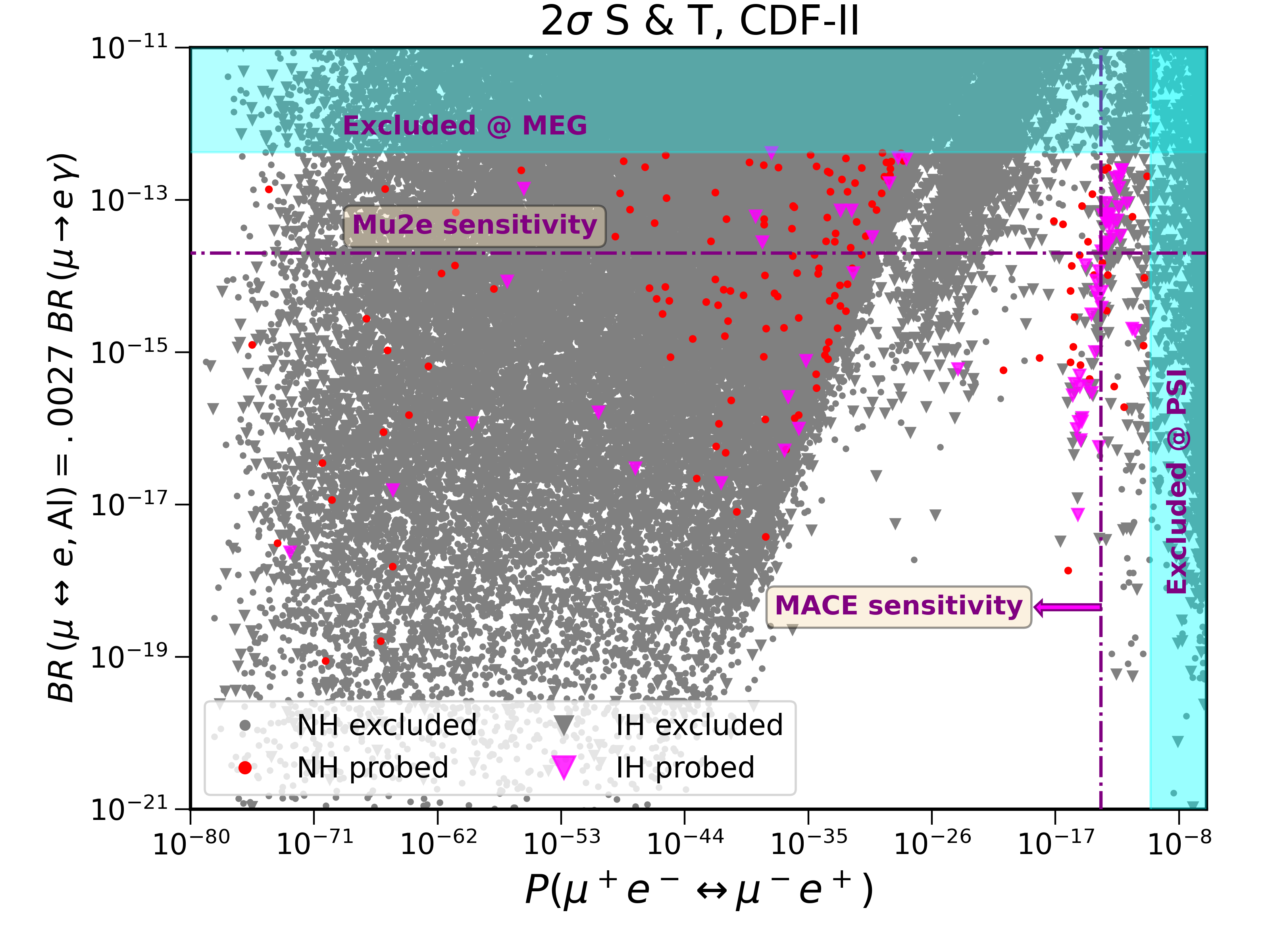}
    \caption{LFV observables muonium--antimuonium conversion and $\mu \to e \gamma$ for normal ($\bullet$) and inverted ($\blacktriangledown$) $M^\nu$ ordering.}
    \label{fig:muonium}
\end{figure}

As can already seen by eye from Fig.~\ref{fig:prediction}, explaining or omitting CDF does not lead to any qualitative differences in our results, in particular with respect to LFV predictions. It is $a_\mu$ that enforces the flavor structure, CDF only requires a particular mass-splitting within the scalar doublet, which does not have a large impact on other observables.
We also find that the neutral scalar that is responsible for the dominant contribution to $(g-2)_\mu$ has a wide mass range of $\unit[20]{GeV}$ to $\unit[3.3]{TeV}$.     
Which process in particular dominates varies from point to point, but $\mu \to e \gamma$/$\mu$-to-$e$ conversion, $\tau \to \ell \mu^+ \mu^-$, and electric dipole moments are generically important. Muonium--antimuonium conversion also probes an important part of the parameter space, see Fig.~\ref{fig:muonium}.

Almost the entire parameter space that can resolve $a_\mu$ -- whether or not we also resolve the CDF anomaly is not relevant -- can be probed with near-future LFV experiments, notably Mu2e and Belle II.
Only a few regions in parameter space remain out of immediate reach, indicated by blue and green points in Fig.~\ref{fig:prediction}. Those points are all small perturations around the Yukawa structures of Eq.~\eqref{eq:muontx1} or \eqref{eq:muontx2}.
Despite leading to suppressed LFV rates, these textures are nevertheless predictive in that they require either $M^\nu_{ee}=0$ or $M^\nu_{\mu\mu}=0$. The former can only be realized for normal hierarchy, gives vanishing $0\nu\beta\beta$, and $\sum_i m_i \in [60,65 ]\,\unit{meV}$. The latter allows for both normal and inverted ordering and predicts rather large values for the lightest neutrino mass, the sum of neutrino masses, and the effective Majorana neutrino mass $\langle m_{\beta\beta}\rangle = |M^\nu_{ee}|$ relevant for $0\nu\beta\beta$, see Tab.~\ref{tab:texture_zeros}.
In fact, limits on $0\nu\beta\beta$ from KamLAND-Zen~\cite{KamLAND-Zen:2022tow} and GERDA~\cite{GERDA:2020xhi} already reach the predicted lower bound, depending on the assumed nuclear matrix elements~\cite{Pompa:2023jxc}. Cosmology constraints on $\sum_i m_i$~\cite{Planck:2018vyg,DiValentino:2021hoh} also reach the predicted lower value for $M^\nu_{\mu\mu}=0$, depending on the combined data sets.
Future improvements on both fronts can probe these predictions unequivocally. For tests of these texture zeros at DUNE using the atmospheric mixing angle and Dirac CP phase, see Ref.~\cite{Bora:2016ygl}.

\section{Conclusion}
\label{sec:conclusion}

The Zee model is one of the oldest and simplest mechanisms for neutrino masses, which occur at one-loop level. The required scalars not only generate Majorana $M^\nu$, but also have couplings to charged leptons that can lead to LFV unsuppressed by $M^\nu$. Here, we have shown that the Zee model can resolve the long-standing anomaly of the muon's magnetic moment, and also the even more significant CDF $W$-mass anomaly. The former requires a particular Yukawa structure and relatively light scalars, which in general leads to dangerously fast LFV processes. While current constraints can be satisfied, the simultaneous explanation of $(g-2)_\mu$ and neutrino masses predicts almost unavoidably LFV in reach of currently-running/near-future experiments such as Belle-II and Mu2e.
We have identified the few finetuned textures that can evade even future LFV limits and shown that they require neutrino-mass texture zeros, either $M^\nu_{ee}=0$ or $M^\nu_{\mu\mu}=0$, which are testable in a complementary way in the neutrino sector.
Overall, we hence find that the Zee-model explanation of $(g-2)_\mu$ is entirely testable/falsifiable. Additionally explaining the CDF anomaly does not modify this conclusion.

\section*{Acknowledgements}

This work was supported in part by the National Science Foundation under Grant PHY-2210428. We acknowledge Research Computing at The University of Virginia for providing computational resources that have contributed to the results reported within this publication.

\appendix

\section*{Appendix: \texorpdfstring{$B_2$}{B2} texture zero}

In this appendix we briefly discuss the $Y_{B_2}$ texture from Eq.~\eqref{eq:TXB2} mentioned in Sec.~\ref{sec:limiting_cases}, which  evades all $\Delta L_\alpha = 1$ LFV. 
TX-$Y_{B_2}$ give rise to $M^\nu_{e\mu}=M^\nu_{\tau\tau}=0$, with predictions for both normal and inverted neutrino-mass hierarchy~\cite{Alcaide:2018vni}. 
Using Eq.~\eqref{eq:Yuk} to solve for this texture leads to the following relations:
\begin{align}
\begin{split}
    Y_{\mu e} &= \frac{M^\nu_{ee}}{2 \kappa m_\mu f_{e\mu}}, 
    \hspace{5mm} Y_{e \mu} = -\frac{M^\nu_{\mu\mu}}{2 \kappa m_e f_{e\mu}} ,  \\
   Y_{\tau \tau} &= \frac{1}{m_\tau \kappa f_{\mu\tau}} \left( 
  M^\nu_{\mu\tau} - \frac{f_{e\tau}}{2 f_{e\mu}} M^\nu_{\mu\mu} \right) .
  \end{split}
  \label{eq:texYe2}
\end{align}
Using the $B_2$ predictions for the currently-unknown neutrino parameters gives an almost real ratio~\cite{Kitabayashi:2015jdj}
\begin{align}
M^\nu_{\mu\mu}/M^\nu_{ee}\simeq 1-\tan^2 \theta_{23}\simeq -1/3\,,
\end{align}
and hence $Y_{e\mu}/Y_{\mu e}\sim 70$.
  The Yukawa couplings $Y_{e\mu}$ and $Y_{\mu e}$ give rise to $(g-2)_\mu$, but also  $(g-2)_e$, $e$EDM, and muonium--antimuonium oscillation.
  We can adjust the phase of $f_{e\mu}$ to render $Y_{\mu e}$ real, which then makes $Y_{e\mu }$ approximately real as well, evading EDM constraints.
Texture~\eqref{eq:texYe2} requires  $\sqrt{|Y_{\mu e}|^2 + |Y_{e\mu}|^2} \approx 1.47\ (m_H/{\rm 100\ GeV})$ to explain $a_\mu$. Inserting these couplings into the muonium--antimuonium probability of Eq.~\eqref{eq:probmm} gives values far in excess of the current limit. Even finetuning $m_A/m_H$ to suppress this observable is not nearly sufficient, thus ruling out this simple texture.

\bibliographystyle{utcaps_mod}
\bibliography{BIB}

\end{document}